\documentclass[a4paper,11pt]{article}
\usepackage{pos}

\newcommand{\centeredgraphics}[2][]{\vcenter{\hbox{\includegraphics[#1]{#2}}}}
\newcommand{\mpl}{m_{\text{Pl}}}
\renewcommand{\logo}{\relax}

\DeclareMathOperator{\diag}{diag}

\definecolor{darkgreen}{rgb}{0,0.5,0}

\title{Gravity in binary systems at the fifth and sixth post-Newtonian order}
\ShortTitle{Gravity in binary systems at the fifth and sixth post-Newtonian order}

\author[a]{J. Blümlein}
\author*[a]{A. Maier}
\author[a]{P. Marquard}
\author[b]{G. Schäfer}

\affiliation[a]{Deutsches Elektronen-Synchrotron DESY,\\
  Platanenallee 6, 15738 Zeuthen, Germany}

\affiliation[b]{Theoretisch-Physikalisches Institut, Friedrich-Schiller-Universität,\\
Max-Wien-Platz 1, D-07743 Jena, Germany}

\emailAdd{Johannes.Bluemlein@desy.de}
\emailAdd{andreas.martin.maier@desy.de}
\emailAdd{peter.marquard@desy.de}
\emailAdd{gerhard.schaefer@uni-jena.de}

\abstract{%
  Binary sources of gravitational waves in the early inspiral phase
  are accurately described by a post-Newtonian expansion in small
  velocity and weak interaction. We compute the conservative dynamics to
  fifth and partial sixth order using a non-relativistic effective field
  theory. We give predictions for central observables and determine the
  required coefficients for the construction of an Effective One-Body
  Hamiltonian, extending the applicability of our results to the late
  inspiral and merger phases.
}

\FullConference{%
  Loops and Legs in Quantum Field Theory - LL2022,\\
  25-30 April, 2022\\
  Ettal, Germany
}

\renewcommand{\hookAfterAbstract}{%
\par\bigskip
DESY-22-131, DO-TH 22/19, SAGEX--22-29
}

\begin{document}
\maketitle

\section{Introduction}

Accurate wave form templates are indispensible to determine the
properties of binary black hole or neutron star systems at
gravitational wave observatories. The construction of such templates
requires a detailed understanding of all phases of binary mergers.

Here, we mainly focus on binary systems of point-like objects without
spin in the early inspiral phase, characterised by a large separation
and small velocities. As a first approximation, the system can be considered
Newtonian with energy
\begin{equation}
  \label{eq:E_N}
  E_N = \mu \left(\frac{\mathrm{v}^2}{2} - \frac{G M}{r} \right)
\end{equation}
in the centre-of-mass frame. $M = m_1 + m_2$ is the sum of the
constituent masses $m_1, m_2$, $\mu = \frac{m_1m_2}{M}$ the reduced
mass, $\mathrm{v}$ the relative velocity, $G$ Newton's constant, and $r$ the
orbital separation. To obtain relativistic corrections, we perform a
post-Newtonian (PN) expansion in $\mathrm{v} \sim \sqrt{\frac{GM}{r}} \ll 1$,
where the correlation between velocity and interaction strength
follows directly from the virial theorem for equation~(\ref{eq:E_N}).

Using a non-relativistic effective theory, the dynamics of compact
binary systems in the post-Newtonian expansion can be computed in
terms of Feynman diagrams. We briefly outline this approach and
present results for the binding energy and the periastron advance at
the fifth post-Newtonian order. We also derive the
local-in-time contributions to the scattering angle
for unbound systems and determine the coefficients of an Effective
One-Body (EOB) Hamiltonian, which allows to predict the dynamics far
beyond the validity range of the post-Newtonian expansion.

\section{Effective Field Theory Framework}
\label{sec:EFT}

\subsection{General Relativity}
\label{sec:GR}

Our starting point is the $d$-dimensional general relativity action in
harmonic gauge, coupled to point-like matter:
\begin{equation}
  \label{eq:S_GR}
  S_{\text{GR}}[g^{\mu\nu}] = \frac{1}{16 G \pi} \int d^dx \sqrt{-g} \left(R - \frac{1}{2} \Gamma_\mu \Gamma^\mu\right) + \sum_{a=1}^2 m_a \int d\tau_a.
\end{equation}
We use natural units, $c = \hbar = 1$, keeping the dependence on
Newton's constant $G$ explicit.
The only dynamic field is the metric $g^{\mu\nu}$ with determinant
$g$. In the limit of flat four-dimensional spacetime, $g^{\mu\nu} \to \eta^{\mu\nu}, d \to 4$,
we choose the signature $\eta^{\mu\nu} = \diag(-1,1,1,1)$. $R$ is the Ricci scalar and $\Gamma^\mu =
g^{\alpha\beta}\Gamma^\mu_{\alpha\beta}$ a contracted Christoffel
symbol. $\tau_a$ is the proper time of the compact object $a$.

Before constructing an effective field theory, we perform a post-Newtonian expansion of the general relativity action. We employ the convenient parametrisation~\cite{Kol:2007rx}
\begin{equation}
  \label{eq:g_KS}
  g^{\mu\nu} = e^{2\textcolor{blue}{\tilde{\phi}}}
  \begin{pmatrix}
    -1 & \textcolor{red}{\tilde{A}_j}\\
    \textcolor{red}{\tilde{A}_i} &
    e^{-c_d \textcolor{blue}{\tilde{\phi}}}(\delta_{ij} + \textcolor{darkgreen}{\tilde{\sigma}_{ij}})-\textcolor{red}{\tilde{A}_i \tilde{A}_j}
  \end{pmatrix}\,,\qquad c_d = 2 \frac{d-2}{d-3}
\end{equation}
with $\textcolor{blue}{\tilde{\phi}} \sim \textcolor{red}{\tilde{A}}
\sim \textcolor{darkgreen}{\tilde{\sigma}} \sim \sqrt{\frac{GM}{r}}$.

Introducing the Planck mass $\mpl = \frac{1}{\sqrt{32 \pi G}}$ and
rescaling $(\textcolor{blue}{\tilde{\phi}},
\textcolor{red}{\tilde{A}}, \textcolor{darkgreen}{\tilde{\sigma}}) =
\frac{1}{\mpl}(\textcolor{blue}{\phi}, \textcolor{red}{A},
\textcolor{darkgreen}{\sigma})$ we arrive at
\begin{equation}
  \label{eq:S_GR_exp}
\begin{split}
  &S_{\text{GR}}[\textcolor{blue}{\phi}, \textcolor{red}{A},
  \textcolor{darkgreen}{\sigma}] =
 \int dt\ \overbrace{\sum_{a=1}^2\left(m_a + \frac{1}{2} m_a \mathrm{v}_a^2 + \dots\right)}^{M+T}\\
  &\qquad+  \int dt\ \sum_{a=1}^2\frac{m_a}{\mpl}\left(-\textcolor{blue}{\phi}(t, \vec{x}_a) + (\mathrm{v}_a)_i\textcolor{red}{A_i}(t, \vec{x}_a) + (\mathrm{v}_a)_i(\mathrm{v}_a)_j\textcolor{darkgreen}{\sigma_{ij}}(t, \vec{x}_a) - \frac{1}{2\mpl} \textcolor{blue}{\phi}(t, \vec{x}_a)^2 + \dots\right) \\
  &\qquad+ \int d^dx\ \left[- c_d \bigl(\partial_\mu \textcolor{blue}{\phi}(x)\bigr)^2 + \bigl(\partial_\mu \textcolor{red}{A_i}(x)\bigr)^2 + \frac{1}{4}\bigl(\partial_\mu \textcolor{darkgreen}{\sigma_{ii}}(x)\bigr)^2 - \frac{1}{2}\bigl(\partial_\mu \textcolor{darkgreen}{\sigma_{ij}}(x)\bigr)^2 + \dots\right],
\end{split}
\end{equation}
where the ellipses indicate higher-order terms omitted for
brevity here. The first line of
equation~(\ref{eq:S_GR_exp}) describes the
matter dynamics, the second line the interaction between matter and
gravitational field, and the third line the bulk field dynamics.

To achieve a consistent expansion, we have to decompose each field
into hard, soft, potential, and radiation (or ultrasoft)
modes~\cite{Beneke:1997zp}. However, hard and soft modes only contribute through quantum
corrections and can be safely neglected. The decomposition into modes therefore reads
\begin{align}
  \label{mode_dec}
  (\textcolor{blue}{\phi}, \textcolor{red}{A},
\textcolor{darkgreen}{\sigma}) ={}& (\textcolor{blue}{\phi_{\text{pot}}}, \textcolor{red}{A_{\text{pot}}},
\textcolor{darkgreen}{\sigma_{\text{pot}}}) + (\textcolor{blue}{\phi_{\text{rad}}}, \textcolor{red}{A_{\text{rad}}},
 \textcolor{darkgreen}{\sigma_{\text{rad}}}),\\
  \label{scaling_pot_spatial}
  \partial_i (\textcolor{blue}{\phi_{\text{pot}}}, \textcolor{red}{A_{\text{pot}}},
\textcolor{darkgreen}{\sigma_{\text{pot}}}) \sim{}& \frac{1}{r} (\textcolor{blue}{\phi_{\text{pot}}}, \textcolor{red}{A_{\text{pot}}},
 \textcolor{darkgreen}{\sigma_{\text{pot}}}),\\
    \label{scaling_pot_time}
  \partial_0 (\textcolor{blue}{\phi_{\text{pot}}}, \textcolor{red}{A_{\text{pot}}},
\textcolor{darkgreen}{\sigma_{\text{pot}}}) \sim{}& \frac{\mathrm{v}}{r} (\textcolor{blue}{\phi_{\text{pot}}}, \textcolor{red}{A_{\text{pot}}},
 \textcolor{darkgreen}{\sigma_{\text{pot}}}),\\
      \label{scaling_rad}
  \partial_\mu (\textcolor{blue}{\phi_{\text{rad}}}, \textcolor{red}{A_{\text{rad}}},
\textcolor{darkgreen}{\sigma_{\text{rad}}}) \sim{}& \frac{\mathrm{v}}{r} (\textcolor{blue}{\phi_{\text{rad}}}, \textcolor{red}{A_{\text{rad}}},
  \textcolor{darkgreen}{\sigma_{\text{rad}}}).
\end{align}

\subsection{Non-Relativistic General Relativity}
\label{sec:NRGR}

We now seek to construct a physically equivalent effective field
theory without potential modes. This effective theory, termed
Non-Relativistic General Relativity (NRGR)~\cite{Goldberger:2004jt}, has an action of the
form
\begin{equation}
  \label{eq:S_NRGR}
  S_{\text{NRGR}}[\textcolor{blue}{\phi_{\text{rad}}}, \textcolor{red}{A_{\text{rad}}},
\textcolor{darkgreen}{\sigma_{\text{rad}}}] = S_{\text{matter}}  + S_{\text{mult}} + S_{\text{bulk, rad}},
\end{equation}
where
\begin{equation}
  \label{eq:S_matter}
  S_{\text{matter}} = \int dt\ (M + T - V_{\text{NZ}})
\end{equation}
describes the matter and its interactions via the classical near-zone potential $V_{\text{NZ}}$,
$S_{\text{mult}}$ the multipole-expanded interactions between matter and radiation modes,
and $S_{\text{bulk, rad}}$ the bulk dynamics of the radiation
modes. Indeed, $S_{\text{bulk, rad}}$ is identical to the general
relativity bulk action after setting the potential modes to
zero. Demanding equivalence to general relativity within the
post-Newtonian expansion leads to the matching condition
\begin{equation}
  \label{eq:V_NZ_match}
  V_{\text{NZ}} = i \frac{d}{dt}\log\left[
    \frac{1}{Z}
    \int \mathcal{D}\textcolor{blue}{\phi_{\text{pot}}}\mathcal{D}\textcolor{red}{A_{\text{pot}}}\mathcal{D}\textcolor{darkgreen}{\sigma_{\text{pot}}}\ %
    e^{i \left[S_{\text{GR}} - \int dt (M+T)\right]_{\textcolor{blue}{\phi_{\text{rad}}}= \textcolor{red}{A_{\text{rad}}}=
\textcolor{darkgreen}{\sigma_{\text{rad}}} = 0}}\right].
\end{equation}
After inserting the action from equation~(\ref{eq:S_GR_exp}) we obtain the following diagrammatic expansion
\begin{equation}
  \label{eq:V_NZ_dias}
  V_{\text{NZ}} = i \frac{d}{dt} \Biggl[\ \underbrace{\centeredgraphics{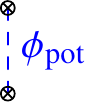}}_{0\text{PN}}
  \ +\ \underbrace{
    \centeredgraphics{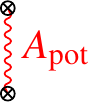}
     \ +\ \frac{1}{2}\times\centeredgraphics{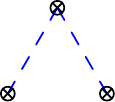}
     \ +\ \frac{1}{2}\times\centeredgraphics{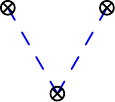}
     }_{1\text{PN}}
    \ +\ (2\text{PN})
    \ \Biggr]\\
\end{equation}
Crossed circles indicate classical sources, i.e. interactions with
compact object 1 at the top and object 2 at the bottom. Inserting the
Feynman rules leads to Fourier transforms of massless propagator-type
Feynman diagrams in $d-1$ space dimensions. Generally, Diagrams with
$L+2$ sources in equation~(\ref{eq:V_NZ_dias}) first contribute at $L$PN
order and correspond to $L$-loop propagator diagrams.\footnote{Certain
classes of diagrams can be factorised into products with a smaller
number of loops in each factor, see~\cite{Foffa:2019hrb,Foffa:2020nqe}.}

We have computed the full 5PN contributions with up to five
loops~\cite{Blumlein:2019zku,Blumlein:2020pog,Blumlein:2020pyo} and
6PN contributions with up to three
loops~\cite{Blumlein:2020znm,Blumlein:2021txj} to
$V_{\text{NZ}}$. Diagrams are generated with
\texttt{QGRAF}~\cite{Nogueira:1991ex}, manipulated with the help of
\texttt{FORM}~\cite{Vermaseren:2000nd,Tentyukov:2007mu}, and reduced
to master integrals using a custom implementation~\cite{crusher} of
Laporta's algorithm~\cite{Laporta:2001dd} for integration-by-parts
reduction~\cite{Chetyrkin:1981qh}. Only one of the master integrals
cannot be calculated with elementary methods; it was first computed
in~\cite{Lee:2015eva}. The result for $V_{\text{NZ}}$ initially
contains higher-order time derivatives, which we systematically
eliminate through integration by parts, multiple-zero insertions, and
coordinate shifts. We find full agreement with independent partial PN
calculations~\cite{Foffa:2019hrb,Foffa:2020nqe} and a post-Minkowskian
expansion to fourth
order~\cite{Bern:2019nnu,Bern:2021dqo,Kalin:2020fhe,Dlapa:2021npj}.

\subsubsection{Post-Newtonian Mechanics}
\label{sec:PN_mech}

Finally, we integrate out the remaining radiation modes, projecting
onto the conservative contribution to the action. We arrive at a post-Newtonian action of the form
\begin{equation}
  \label{eq:S_PN}
  S_{\text{PN}} = \int dt\ L[\vec{r}, \vec{\mathrm{v}}] = \int dt\ (M + T - V_{\text{NZ}} + L_{\text{FZ}}),
\end{equation}
where the far-zone Lagrangian $L_{\text{FZ}}$ is obtained from the
NRGR action, equation~(\ref{eq:S_NRGR}), through the matching
condition
\begin{equation}
  \label{eq:L_FZ_matching}
  L_{\text{FZ}} = -i \frac{d}{dt} \log\left[
    \frac{1}{\tilde{Z}}
    \int \mathcal{D}\textcolor{blue}{\phi_{\text{rad}}}\mathcal{D}\textcolor{red}{A_{\text{rad}}}\mathcal{D}\textcolor{darkgreen}{\sigma_{\text{rad}}}\ %
    e^{i S_{\text{rad}}}\right]_{\text{cons}},
\end{equation}
where $S_{\text{rad}} = S_{\text{mult}} + S_{\text{bulk, rad}}$. The
subscript indicates projection onto the conservative
contribution. $S_{\text{mult}}$ describes the interactions between
matter and radiation modes in NRGR. Since the wave length of the
radiation modes is parametrically larger than the orbital separation
and the wave length of the potential modes, c.f.\
equations~\eqref{scaling_pot_spatial} and \eqref{scaling_rad}, this
contribution to the action has to be multipole-expanded. In the
centre-of-mass frame one obtains in $d$ dimensions~\cite{Ross:2012fc,Almeida:2021xwn}
\begin{equation*}
  \label{eq:S_mult}
      \begin{split}
      S_{\text{mixed}} = \int dt\ &\left[\frac{1}{2} {E} \delta g_{00}(t, \vec{0}) - \frac{1}{2} {L^{i|j}} \partial_j \delta g_{0i}(t, \vec{0}) - \frac{1}{2} {Q^{ij}} R_{0i0j}(t, \vec{0}) \right.\\
      &  \left.- \frac{1}{3} {J^{i|jk}} R_{0jki}(t, \vec{0})
      - \frac{1}{6} {O^{ijk}} \partial_k R_{0i0j}(t, \vec{0}) + \dots \right],
  \end{split}
\end{equation*}
where $R^\mu_{\phantom{\mu}\nu\rho\sigma}$ is the Riemann tensor,
expanded in powers of $\delta g_{\mu\nu} = g_{\mu\nu} -
\eta_{\mu\nu}$. We again use the parametrisation defined in equation~(\ref{eq:g_KS}) for the metric, replacing $(\textcolor{blue}{\phi}, \textcolor{red}{A},
\textcolor{darkgreen}{\sigma}) \to (\textcolor{blue}{\phi_{\text{rad}}}, \textcolor{red}{A_{\text{rad}}},
\textcolor{darkgreen}{\sigma_{\text{rad}}})$. We distinguish between mass-type (``electric'')
multipole moments $E, P^i = 0, Q^{ij}, O^{ijk}$ which are symmetric
under permutation of indices, and current-type (``magnetic'')
multipole moments $L^{i|j}, J^{i|jk}$ which are antisymmetric under
index exchanges crossing the vertical bar. In principle, further
evanescent multipole moments arise for $d\neq 4$. However, they do not
contribute at 5PN order.

To disentangle conservative and dissipative effects we employ the
closed-time-path (or in-in) formalism, replacing
\begin{equation}
  \label{eq:in-in}
 S_{\text{rad}}[\textcolor{blue}{\phi_{\text{rad}}}, \textcolor{red}{A_{\text{rad}}},
 \textcolor{darkgreen}{\sigma_{\text{rad}}}, \vec{r}, \vec{\mathrm{v}}]
 \to
 S_{\text{rad}}[\textcolor{blue}{\phi_{\text{rad}, 1}}, \textcolor{red}{A_{\text{rad}, 1}},
 \textcolor{darkgreen}{\sigma_{\text{rad}, 1}}, \vec{r}_1, \vec{\mathrm{v}}_1]
 - S_{\text{rad}}[\textcolor{blue}{\phi_{\text{rad}, 2}}, \textcolor{red}{A_{\text{rad}, 2}},
\textcolor{darkgreen}{\sigma_{\text{rad}, 2}}, \vec{r}_2, \vec{\mathrm{v}}_2]
\end{equation}
in the matching equation~(\ref{eq:L_FZ_matching}),
cf.~\cite{Blumlein:2021txe}. Note that the
subscripts here indicate the time path and are unrelated to the
indices associated with the compact objects.

From equation~(\ref{eq:L_FZ_matching}) we obtain
\begin{equation}
  \label{eq:L_FZ_dias}
 L_{\text{FZ}} = -i \frac{d}{dt} \left[\ \centeredgraphics{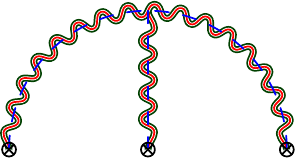}\ +\ \centeredgraphics{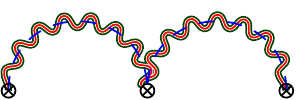}\ \right]_{\text{cons}},
\end{equation}
where the lines denote retarded or advanced propagators of
$\textcolor{blue}{\phi_{\text{rad}}},
\textcolor{red}{A_{\text{rad}}}$, or
$\textcolor{darkgreen}{\sigma_{\text{rad}}}$ and the crossed circles
are interactions with the multipole moments of the binary system. At
4PN order, only the combination of one coupling to the energy $E$ and
two couplings to the mass quadrupole $Q_{ij}$ contributes, while at
5PN order all of the multipole moments in equation~\eqref{eq:S_mult}
appear. This implies that $d$-dimensional expressions including 1PN
corrections are required for $E$ and $Q$~\cite{Marchand:2020fpt}, whereas the Newtonian level
is sufficient for the remaining multipole moments. Retaining only
terms that are linear in $\vec{r}_- = \frac{1}{\sqrt{2}}(\vec{r}_1 - \vec{r}_2)$, the
result has the form
\begin{equation}
  \label{eq:L_FZ_12}
 L_{\text{FZ}} = -i \frac{d}{dt} \bigg[S_{\text{hom}}[\vec{r}_1, \vec{\mathrm{v}}_1] - S_{\text{hom}}[\vec{r}_2, \vec{\mathrm{v}}_2] + S_{\text{inhom}}[\vec{r}_1,\vec{r}_2, \vec{\mathrm{v}}_1, \vec{\mathrm{v}}_2] \bigg]_{\text{cons}}.
\end{equation}
We identify $S_{\text{hom}}$ with the conservative
contribution~\cite{Galley:2015kus}. Note that $L_{\text{FZ}}$ can be
subdivided into a part $L_{\text{FZ, loc}}$ describing local-in-time,
i.e.\ instantaneous, interactions and a non-local part $L_{\text{FZ,
nl}}$ containing a time integral that cannot be evaluated in closed
form without specifying a trajectory.

\section{Observables}
\label{sec:observables}

After a Legendre transform, we obtain a Hamiltonian, from which we
predict physical observables. We briefly review our main results; for
a detailed account see~\cite{Blumlein:2021txe}.

\begin{figure}[htb]
  \centering
\includegraphics[height=45mm]{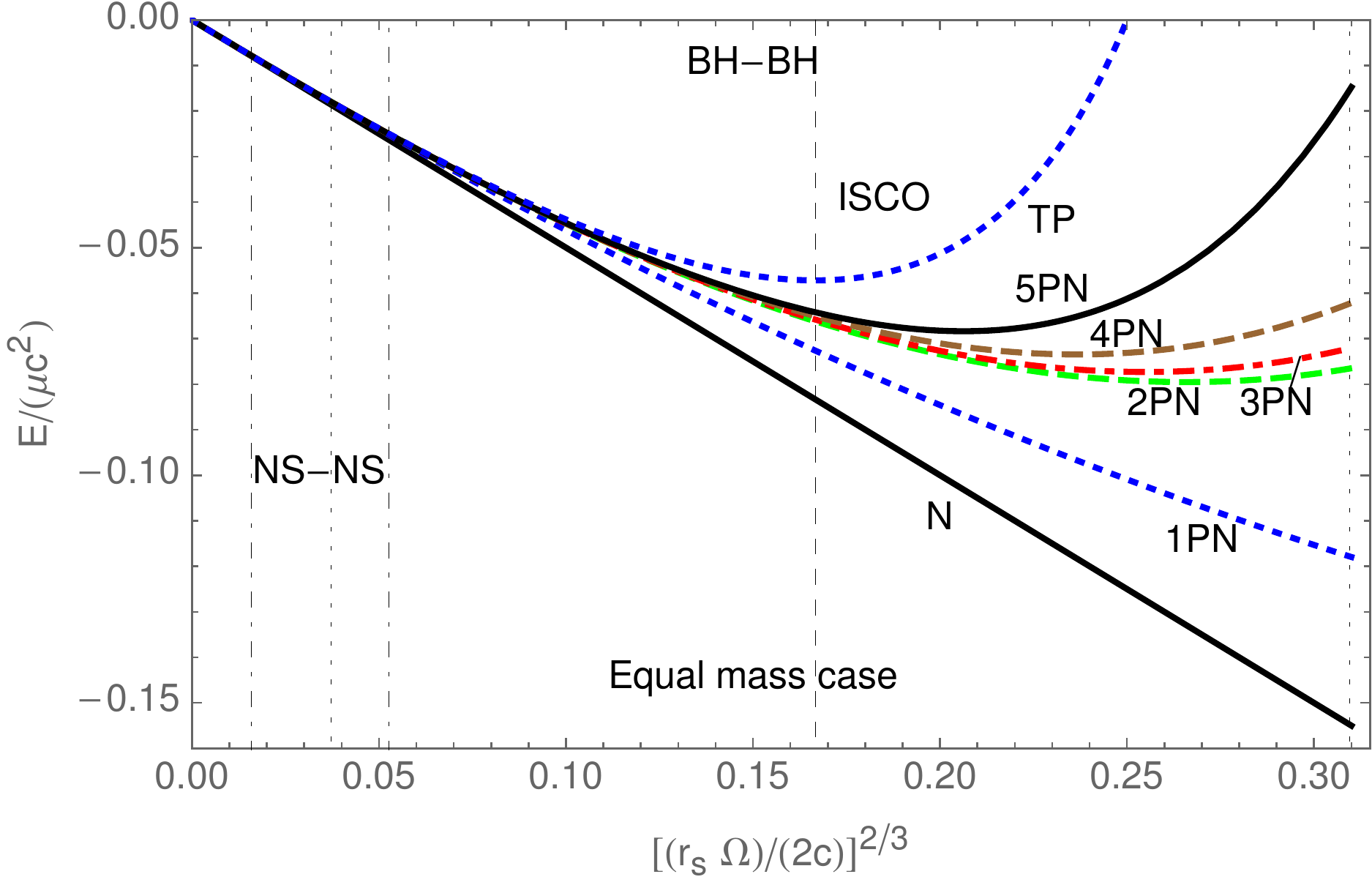}\qquad \includegraphics[height=45mm]{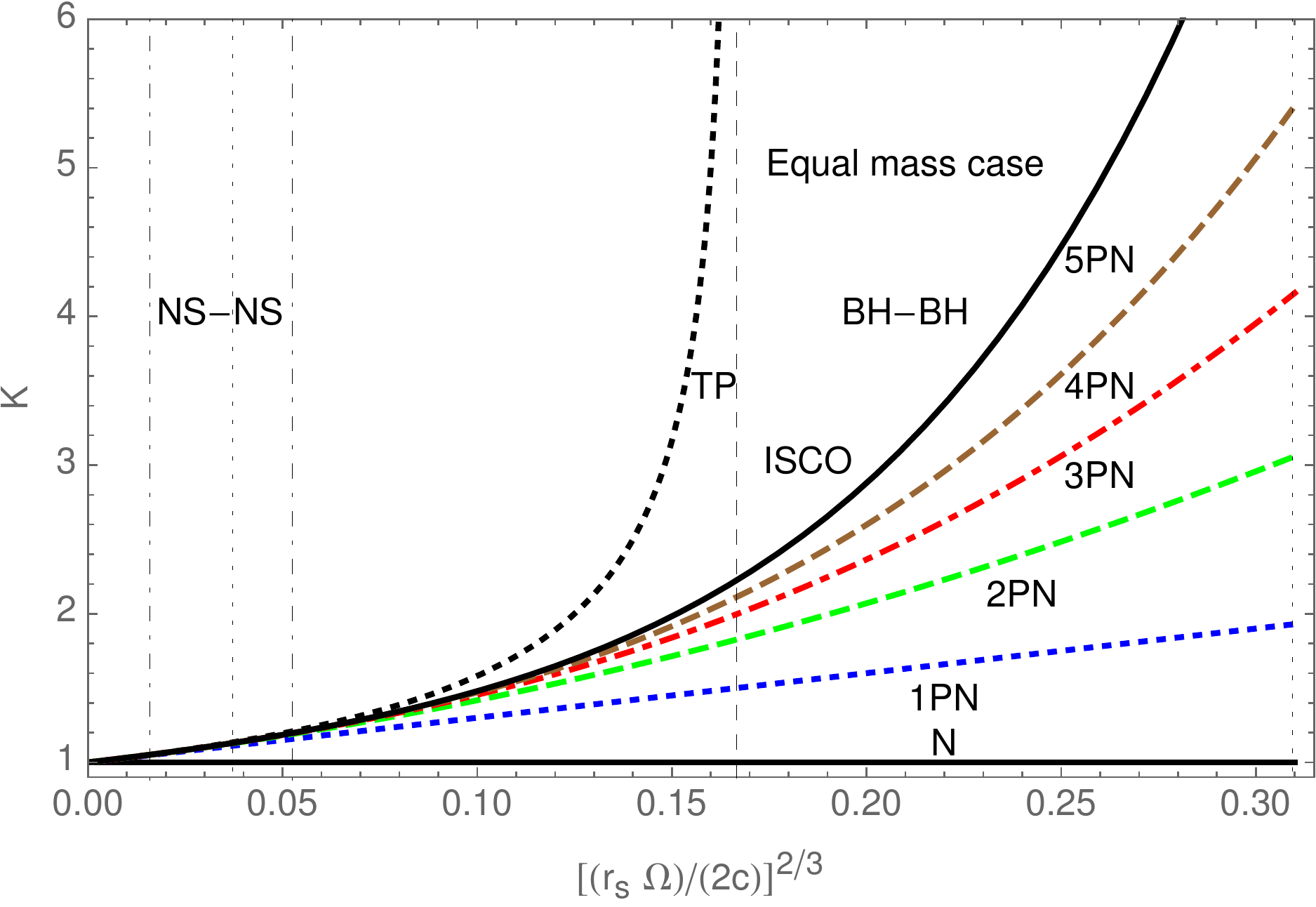}
\caption{
  Energy (left) and Periastron advance (right) as functions of
  $x = (G M \Omega)^{\frac{2}{3}}$ in the quasi-circular case for
  equal masses. The uppermost dashed line denotes the exact result in
  the test particle limit, where the innermost stable circular orbit
  (ISCO) is marked by a dashed vertical line. The solid black lines
  and curves in between show predictions from successive
  post-Newtonian orders. The dotted and dash-dotted
  vertical lines illustrate the frequency ranges for binary black-hole
  and neutron star systems covered by current experiments. Figure
  reproduced from~\cite{Blumlein:2021txe}.}
  \label{fig:E_K}
\end{figure}

First, we consider an elliptic orbit in the limit of vanishing
eccentricity. The binding energy is given by
\begin{equation}
  \label{eq:E}
  E^{\text{circ}} = E^{\text{circ}}_{\text{loc}} + E^{\text{circ}}_{\text{nl}},
\end{equation}
where $E^{\text{circ}}_{\text{loc}}$ and $E^{\text{circ}}_{\text{nl}}$
denote the contributions originating from local and non-local in time
Hamiltonian parts, respectively. Expressed in terms of the reduced
angular momentum $j = \frac{J}{GM}$ and $\nu = \frac{\mu}{M}$ they read
    \begin{equation}
    \begin{split}
\frac{E_{\text{loc}}^{\text{circ}}(j)}{\mu}
={}& -\frac{1}{2j^2}
+\left(-\frac{\nu }{8}-\frac{9}{8}\right)\frac{1}{j^4}
+\left(-\frac{\nu ^2}{16}+\frac{7 \nu }{16}-\frac{81}{16}\right)\frac{1}{j^6}
+\Biggl[-\frac{5 \nu ^3}{128}+\frac{5 \nu ^2}{64}+\Biggl(\frac{8833}{384}
-\frac{41 \pi ^2}{64}\Biggr) \nu\\ &
-\frac{3861}{128}
\Biggr] \frac{1}{j^8}
+
\Biggl[-\frac{7 \nu ^4}{256}
+\frac{3 \nu ^3}{128}
+\left(\frac{41 \pi^2}{128}-\frac{8875}{768}\right) \nu^2
+\Biggl(\frac{989911}{3840}
-\frac{6581 \pi ^2}{1024}\Biggr) \nu
\\ &
   -\frac{53703}{256}
\Biggr] \frac{1}{j^{10}}
+
{\Biggl[-\frac{21 \nu^5}{1024}
+\frac{5 \nu ^4}{1024}
+\Biggl(\frac{41 \pi^2}{512}
-\frac{3769}{3072}\Biggr) \nu^3
}
\Biggl(
-\frac{400383799}{403200}
 + \frac{132979 \pi^2}{2048}
     \Biggr) \nu ^2
\\ &
     +\Biggl(
\frac{3747183493}{1612800} - \frac{31547 \pi ^2}{1536}\Biggr) \nu
-\frac{1648269}{1024} \Biggr] \frac{1}{j^{12}}  +\mathcal{O}\left(\frac{1}{j^{14}}\right),
\end{split}
\end{equation}
\begin{equation}
  \begin{split}
\frac{E^\text{circ}_\text{nl}}{\mu} ={}& \nu \Biggl\{
\Biggl[
        -\frac{64}{5} (\ln(j) -\gamma_E)
        +\frac{128}{5} \ln(2) \Biggr] \frac{1}{j^{10}}
+
\Biggl[
        \frac{32}{5}
        +\frac{28484}{105} \ln(2)
        +\frac{243}{14} \ln(3)
    \\ &
        -\frac{15172}{105} (\ln(j) - \gamma_E)
        +\nu  \Biggl(
                \frac{32}{5}+\frac{112}{5} (\ln(j)
- \gamma_E)
                +\frac{912}{35} \ln(2)
- \frac{486}{7} \ln(3) \Biggr)
         \Biggr] \frac{1}{j^{12}}\Biggr\}
\\&+\mathcal{O}\left(\frac{1}{j^{14}}\right)
\end{split}
\end{equation}
up to 5PN order. Similarly, the 5PN periastron advance in this limit is
given by
\begin{align}
  K^{\text{circ}}(j) ={}& K^{\text{circ}}_{\text{loc}}(j) + K^{\text{circ}}_{\text{nl}}(j),\\
K^{\text{circ}}_{\text{loc}}(j)
={}&
1 + 3 \frac{1}{j^2}
+\Biggl(\frac{45}{2}-6\nu\Biggr)\frac{1}{j^4}
+\Biggl[\frac{405}{2}+\Biggl(-202+\frac{123}{32}\pi^2\Biggr)\nu+3\nu^2\Biggr]\frac{1}{j^6}
\notag\\ &
+
\Biggl[\frac{15795}{8}+\left(\frac{185767}{3072}\pi^2
-\frac{105991}{36}\right)\nu
+\Biggl(-\frac{41}{4}\pi^2+\frac{2479}{6}\Biggr)\nu^2\Biggr]\frac{1}{j^8}
  \notag\\ &
+
{\Biggl[\frac{161109}{8}}
+\Biggl(-\frac{18144676}{525} + \frac{488373}{2048}\pi^2\Biggr)\nu
- \Biggl(\frac{15079451}{675}  + \frac{1379075}{1024} \pi^2\Biggr)
     \nu^2
\notag\\ &
+\Biggl(-\frac{1627}{6}+\frac{205}{32}\pi^2\Biggr)\nu^3\Biggr]\frac{1}{j^{10}}
+ O\left(\frac{1}{j^{12}} \right),\\
K^{\text{circ}}_{\text{nl}}(j) ={}& -\frac{64}{10} \nu \Biggl\{\frac{1}{j^8} \Biggl[ - 11
    - \frac{157}{6} (\ln (j) - \gamma_E) + \frac{37}{6} \ln(2) + \frac{729}{16} \ln(3) \Biggr]\notag\\
    &
+ { \frac{1}{j^{10}}
 \Biggl[
-\frac{59723}{336} - \frac{9421}{28} [\ln(j) - \gamma_E] + \frac{7605}{28}  \ln(2)
+ \frac{112995}{224} \ln(3)
}\notag\\ &
 {+ \Biggl(
  \frac{2227}{42}
+ \frac{617}{6} [\ln(j) - \gamma_E]
- \frac{7105}{6} \ln(2)
+ \frac{54675}{112} \ln(3)
\Biggr) \nu
\Biggr]}+ O\left(\frac{1}{j^{12}}\right).
\end{align}
Both observables are shown in figure~\ref{fig:E_K} as functions of $x
= (G M \Omega)^{\frac{2}{3}}$, where $\Omega$ is the angular orbital
frequency with $G M \Omega = \frac{1}{\mu} \frac{dE^{\text{circ}}}{dj}$, see e.g.~\cite{Schafer:2018kuf}.

The frequency, or equivalently $x$, for the innermost stable circular
orbit (ISCO) is found by minimising the energy, $\frac{dE^{\text{circ}}}{dx} = 0$. In the test particle limit $m_2 \to 0$ one finds
\begin{equation}
  \label{eq:ISCO_TP}
  \frac{E^{\text{circ}}_{\text{TP}}}{\mu} = \frac{1-2x}{\sqrt{1-3x}}-1,\qquad x^{\text{ISCO}}_{\text{TP}} = \frac{1}{6}.
\end{equation}
To obtain an $n$PN-accurate expression for the circular binding energy with arbitrary masses, we define
\begin{equation}
  \label{eq:E_PN+TP}
E^{\text{circ}}_{\text{TP} + n\text{PN}\nu} =  E^{\text{circ}}_{\text{TP}} - E^{\text{circ}}_{\text{TP}, n\text{PN}} + E^{\text{circ}}_{n\text{PN}},
\end{equation}
where $E^{\text{circ}}_{\text{TP}, n\text{PN}}$ is
$E^{\text{circ}}_{\text{TP}}$ expanded to $n$PN order and
$E^{\text{circ}}_{n\text{PN}}$ is the $n$PN approximation to the
circular binding energy. For the ISCO frequency we then obtain
\begin{align}
  \label{eq:x_ISCO_1PN}
  x^{\text{ISCO}}_{\text{TP} + 1\text{PN}\nu} \approx{}& 0.16625696117738773332,\\
x^{\text{ISCO}}_{\text{TP} + 2\text{PN}\nu} \approx{}& 0.16918429803993408276,\\
x^{\text{ISCO}}_{\text{TP} + 3\text{PN}\nu} \approx{}& 0.18096727401354580502,\\
x^{\text{ISCO}}_{\text{TP} + 4\text{PN}\nu} \approx{}& 0.18807765161857146375,\\
x^{\text{ISCO}}_{\text{TP} + 5\text{PN}\nu} \approx{}& 0.18788220419180782864.
\end{align}

We compare our results to the local-in-time EOB Hamiltonian~\cite{Bini:2020wpo}, viz.
\begin{align}
  \label{eq:H_EOB}
  H_{\text{EOB}}^{\text{loc}} ={}& \sqrt{A(1+p^2 + (AD-1) p_r^2  + Q)},\\
  \label{eq:A}
  A ={}& 1 + \sum_{k=1}^6 \frac{a_k}{r^k},\\
    \label{eq:D}
  D ={}& 1 + \sum_{k=2}^5 \frac{d_k}{r^k},\\
    \label{eq:Q}
  Q ={}& p_r^4\left[\frac{q_{42}}{r^2}
+ \frac{q_{43}}{r^3} + \frac{q_{44}}{r^4}\right]
+ p_r^6 \left[\frac{q_{62}}{r^2} + \frac{q_{63}}{r^3}\right]
+ p_r^8 \frac{q_{82}}{r^2}
\end{align}
to 5PN, where $p_r$ denotes the radial component of the spatial
momentum and $r = \frac{r_{\text{phys}}}{G M}$ is the rescaled orbital
separation. Comparing to our results for the circular binding energy
and the local periastron advance $K^\text{loc}(E,j)$ beyond the
circular limit we derive all 5PN EOB parameters:
\begin{align}
\label{eq:VAL1}
q_{82} ={}& \frac{6}{7} \nu + \frac{18}{7} \nu^2 + \frac{24}{7} \nu^3 - 6 \nu^4,
\\
\label{eq:VAL2}
q_{63} ={}& \frac{123}{10} \nu - \frac{69}{5} \nu^2 + 116 \nu^3 - 14 \nu^4,
\\
\label{eq:VAL3}
q_{44} ={}&
             \left(\frac{1580641}{3150} - \frac{93031 \pi^2}{1536} \right) \nu
           + \left({
-\frac{3670222}{4725}}
+ \frac{31633 \pi^2}{512} \right) \nu^2
           +\left(640 -\frac{615}{32} \pi ^2\right) \nu ^3,
\\
\label{eq:VAL4}
d_5    ={}&  \left( \frac{331054}{175} - \frac{63707}{512} \pi^2 \right) \nu
          + \left({-\frac{31295104}{4725}} +\frac{306545}{512} \pi^2\right) \nu^2
          + \left(\frac{1069}{3} - \frac{205}{16} \pi^2 \right) \nu^3,
\\
\label{eq:VAL5}
a_6    ={}& \left(- \frac{1026301}{1575} + \frac{246367}{3072} \pi^2\right)\nu +
 \left(
{-\frac{1749043}{1575}}
+ \frac{25911}{256} \pi^2\right) \nu^2 + 4
\nu^3.
\end{align}
All other coefficients in eqs.~\eqref{eq:A}, \eqref{eq:D}, \eqref{eq:Q}
are already determined up to 4PN.
The predictions
for the terms proportional to $\nu^2$ in $d_5$ and $a_6$ have
not been obtained in any other approach so far.

Moving beyond a closed orbit, we compute the PN expansion of the
scattering angle $\chi$. Explicit formulae are given
in~\cite{Blumlein:2021txe}. For the most part, we find agreement with
other determinations~\cite{Bini:2021gat,Bern:2021yeh,Dlapa:2021vgp},
except for a deviation that can be traced back to the rational
contribution of order $\nu^2$ to $q_{44}$.
This numerically small difference, cf.~\cite{Khalil:2022ylj},
is under investigation. Note that it has not yet
been proven that one can perform an analytic continuation of the
Hamiltonian dynamics from scattering to the bound state problem from
5PN onward.

\section*{Acknowledgements}

We thank Z.~Bern, D.~Bini, L.~Blanchet, Th. Damour, S. Foffa,
G.~Kälin, C. Kavanagh, R.~Porto, M.~Ruf, R.~Sturani, and B.~Wardell
for discussions concerning a few aspects related to our earlier
work Ref.~\cite{Blumlein:2020pyo,Blumlein:2021txe}.
This work has been funded in part by EU TMR network SAGEX agreement
No. 764850 (Marie Skłodowska-Curie).

\bibliographystyle{JHEP}
\bibliography{biblio.bib}



\end{document}